\documentclass{aastex631}

\newcommand{\ms}{\text{m\,s}$^{-1}$}

\usepackage{newunicodechar}
\DeclareRobustCommand{\okina}{%
 \raisebox{\dimexpr\fontcharht\font`A-\height}{%
 \scalebox{0.8}{`}%
 }%
}
\newunicodechar{ʻ}{\okina}

\defcitealias{Yee2018}{Y18}
\defcitealias{Basilicata2024}{B24}

\begin{document}

\title{Additional Doppler Monitoring Corroborates HAT-P-11\,c as a Planet}

\author[0000-0001-7961-3907]{Samuel~W.~Yee}
\altaffiliation{51 Pegasi b Fellow}
\affiliation{Center for Astrophysics \textbar \ Harvard \& Smithsonian, 60 Garden St, Cambridge, MA 02138, USA}

\author[0000-0003-0967-2893]{Erik~A.~Petigura}
\affiliation{Department of Physics \& Astronomy, University of California Los Angeles, Los Angeles, CA 90095, USA}

\author[0000-0002-0531-1073]{Howard~Isaacson}
\affiliation{Department of Astronomy,  University of California Berkeley, Berkeley, CA 94720, USA}
\affiliation{University of Southern Queensland, Centre for Astrophysics, West Street, Toowoomba, QLD 4350, Australia}

\author[0000-0001-8638-0320]{Andrew~W.~Howard}
\affiliation{Department of Astronomy, California Institute of Technology, Pasadena, CA 91125, USA}

\author[0000-0002-3199-2888]{Sarah Blunt}
\affiliation{Center for Interdisciplinary Exploration and Research in Astrophysics (CIERA), Northwestern University, 1800 Sherman, Evanston, IL, 60201, USA}

\author[0000-0002-4297-5506]{Paul A.\ Dalba}
\affiliation{Department of Astronomy and Astrophysics, University of California, Santa Cruz, CA 95064, USA}

\author[0000-0002-8958-0683]{Fei Dai}
\affiliation{Institute for Astronomy, University of Hawaiʻi at M\=anoa, 2680 Woodlawn Drive, Honolulu, HI 96822, USA}

\author[0000-0003-3504-5316]{Benjamin~J.~Fulton}
\affiliation{NASA Exoplanet Science Institute / Caltech-IPAC, Pasadena, CA 91125, USA}
\author[0000-0002-8965-3969]{Steven Giacalone}
\altaffiliation{NSF Astronomy and Astrophysics Postdoctoral Fellow}
\affiliation{Department of Astronomy, California Institute of Technology, Pasadena, CA 91125, USA}

\author[0000-0002-7084-0529]{Stephen~R.~Kane}
\affiliation{Department of Earth and Planetary Sciences, University of California, Riverside, CA 92521, USA}

\author[0000-0002-6115-4359]{Molly~Kosiarek}
\affiliation{Department of Astronomy and Astrophysics, University of California, Santa Cruz, CA 95064, USA}

\author[0000-0003-4603-556X]{Teo~Mo\v{c}nik}
\affiliation{Gemini Observatory/NSF's NOIRLab, Hilo, HI 96720, USA}

\author[0000-0002-7670-670X]{Malena Rice}
\affiliation{Department of Astronomy, Yale University, New Haven, CT 06511, USA}


\author[0000-0003-3856-3143]{Ryan~Rubenzahl}
\affiliation{Department of Astronomy, California Institute of Technology, Pasadena, CA 91125, USA}

\author[0000-0003-2657-3889]{Nicholas Saunders}
\altaffiliation{NSF Graduate Research Fellow}
\affiliation{Institute for Astronomy, University of Hawaiʻi at M\=anoa, 2680 Woodlawn Drive, Honolulu, HI 96822, USA}

\author[0000-0003-0298-4667]{Dakotah~Tyler}
\affiliation{Department of Physics \& Astronomy, University of California Los Angeles, Los Angeles, CA 90095, USA}

\author[0000-0002-3725-3058]{Lauren M. Weiss}
\affiliation{Department of Physics and Astronomy, 225 Nieuwland Science Hall, 
University of Notre Dame, Notre Dame, IN 46556, USA}

\author[0000-0002-2696-2406]{Jingwen~Zhang}
\altaffiliation{NASA FINESST Fellow}
\affiliation{Institute for Astronomy, University of Hawaiʻi at M\=anoa, 2680 Woodlawn Drive, Honolulu, HI 96822, USA}

\begin{abstract}

In 2010, Bakos and collaborators discovered a Neptune-sized planet transiting the K-dwarf HAT-P-11 every five days. Later in 2018, Yee and collaborators reported an additional Jovian-mass companion on a nine year orbit based on a decade of Doppler monitoring. The eccentric outer giant HAT-P-11c may be responsible for the peculiar polar orbit of the inner planet HAT-P-11b. However, \citet{Basilicata2024} recently suggested that the HAT-P-11c Doppler signal could be caused by stellar activity. In this research note, we extend the \citet{Yee2018} Doppler time series by six years. The combined dataset spanning 17 years covers nearly two orbits of the outer planet. Importantly, we observe two periastron passages of planet c and do not observe a coherent activity signature. Together with the previously reported astrometric acceleration of HAT-P-11 from Hipparcos and Gaia, we believe there is strong evidence for HAT-P-11c as a \textit{bona fide} planet.
\end{abstract}
\keywords{Exoplanets (498), Radial velocity (1332)} 

\section{Introduction} \label{sec:intro}

HAT-P-11 hosts a transiting hot Neptune \citep{Bakos2010} on a polar orbit \citep[e.g.,][]{Winn2010,Hirano2011}.
A decade of precision radial-velocity (RV) monitoring with the High Resolution Echelle Spectrometer \citep[HIRES;][]{HIRES_Vogt94} on the Keck I telescope revealed HAT-P-11c, a distant Jovian planet in the same system on a $\sim 9$~year, eccentric ($e = 0.6$) orbit \citep[][hereafter Y18]{Yee2018}.

Given that HAT-P-11 is only 38~pc from the Earth, \citetalias{Yee2018} suggested that the astrometric signature of planet c might be detected by the ESA \textit{Gaia} mission.
While epoch astrometry from \textit{Gaia} is yet to be released, HAT-P-11 was found to be accelerating based on changes in its position and proper motion between the \textit{Hipparcos} and \textit{Gaia} surveys \citep{Xuan2020,An2024}.
This astrometric acceleration has been used in conjunction with previously published RV data to infer a significant misalignment between HAT-P-11\,b and c, providing clues to the dynamical history leading to the system's unusual architecture \citep[e.g.,][]{Lu2024}.

One complication is that HAT-P-11 is a moderately active star with a stellar activity cycle of $\gtrsim 10$~years \citep{Morris2017}, comparable to the orbital period of planet c. Care is required to discriminate whether the observed RV variations are due to stellar reflex motion or RV-activity correlations. \citetalias{Yee2018} argued that the $\sim 30$~\ms\ semi-amplitude and eccentric shape of the RV signal, along with a 400 day phase offset between the RV and stellar activity cycles, were indicative that a long-period giant planet is present in the system.
Recently, \citet[][hereafter B24]{Basilicata2024}, revisited the system and attributed the long-period RV signal to stellar activity, citing additional evidence from four series of spectroscopic observations taken during transits of HAT-P-11b between 2019 and 2023, as well as a new discovery that activity-induced RV signals can be offset from chromospheric activity indicators by hundreds of days \citep{Meunier2019}.

In this Note, we present new Keck/HIRES data from continued monitoring of HAT-P-11 between 2018 and 2024, which strengthens the evidence for the planetary scenario.

\section{Corroboration of HAT-P-11c's Planetary Nature}

\begin{figure}
    \centering
    \includegraphics{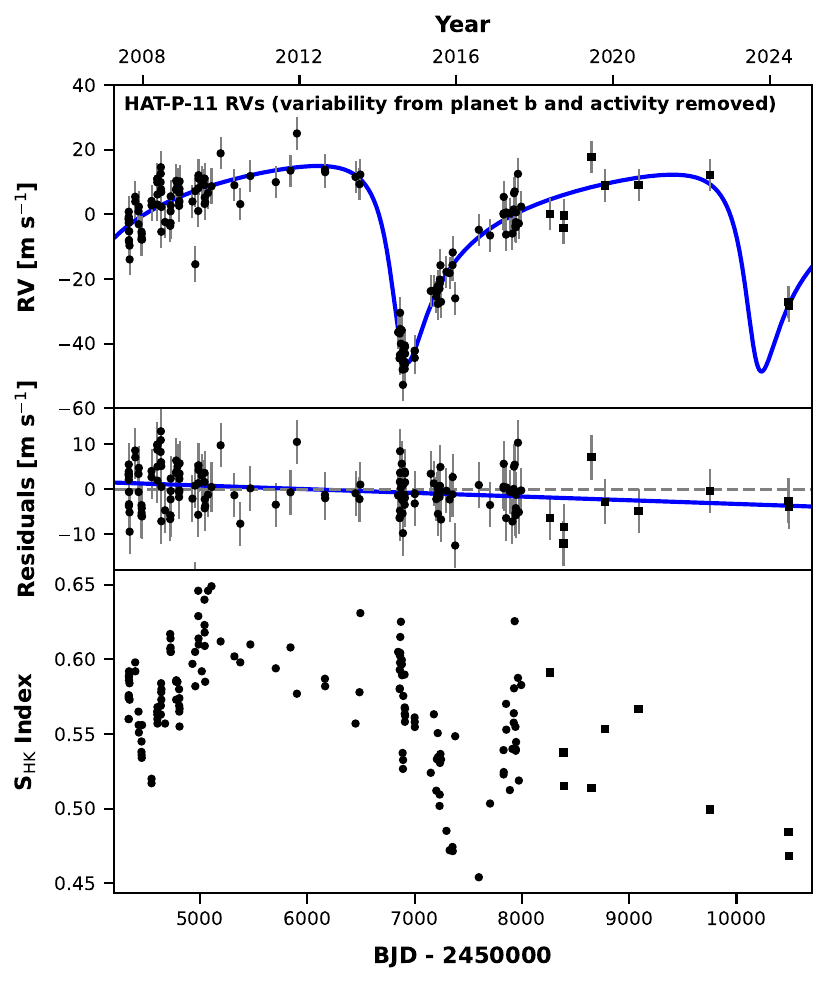}
    \caption{
    \textbf{Top:} RVs of HAT-P-11 after subtracting the best-fitting model for the inner transiting planet b and a linear correlation with the S$_\mathrm{HK}$ index. Data collected after \citetalias{Yee2018} are shown as squares. Error bars include a jitter term added in quadrature to the measurement uncertainties. The blue line shows the best-fit model for the stellar reflex motion due to planet c.
    \textbf{Middle:} Residuals to the best-fit two-planet model.
    \textbf{Bottom:} S$_\mathrm{HK}$ time series for HAT-P-11, which does not show a coherent long-term periodicity.
    The data and best-fit model parameters are available as Data behind the Figure.}
    \label{fig:hat-p-11_rvs}
\end{figure}

We obtained nine new spectra with Keck/HIRES after the publication of \citetalias{Yee2018}. The observations and data reduction were identical to the original publication using the procedures of the California Planet Search \citep[CPS;][]{CPS_Howard2010}.
The full RV and activity time series span 17 years (2007--2024) and are plotted in Figure \ref{fig:hat-p-11_rvs}. The new RV data are consistent with the stellar reflex motion due to a long-period eccentric planet with the same parameters as in \citetalias{Yee2018}.
Meanwhile, the $S_\mathrm{HK}$ values do not track the long-term, large-amplitude RV variations well, with or without a phase offset, suggesting that the stellar activity cycle has a different periodicity from the orbit of planet c and is not the cause of the long-period RV signal.

We modeled the RVs in the same manner as \citetalias{Yee2018}. In brief, we fit a two-Keplerian model and linear RV-$S_\mathrm{HK}$ correlation. \citetalias{Yee2018} found a weak RV-$S_\mathrm{HK}$ correlation accounting for $\lesssim15$~\ms\ peak-to-peak RV variation; the amplitude of this correlation remains consistent with the addition of the new data.
The remaining observed RV variation is 60 \ms\ and is best fit by a Keplerian orbit.

The RVs are consistent with a bimodal distribution for the orbital period of planet c at $\sim3300$ and 3670~days. The two modes correspond to models where periastron passage occurs before or after the most recent HIRES observations in June 2024. We favor the former scenario, as HARPS-N data presented in \citetalias{Basilicata2024} show an RV change of $\approx -50$~\ms\ between 18 June 2020 and 13 June 2023; this large $\Delta(\mathrm{RV})$ is only compatible with the scenario where HAT-P-11c was close to periastron in June 2023.
As such, we plot only this scenario in Figure \ref{fig:hat-p-11_rvs}.
The best-fit properties of HAT-P-11c in this case are $P_c = 3299^{+53}_{-56}$~days, $T_{\mathrm{peri},c} = 2456867^{+15}_{-18}$, $e_c = 0.617 ^{+0.027}_{-0.029}$, $\omega_c = 143.9^{+4.0}_{-3.8}$ deg, $K_c = 30.1 ^{+1.1}_{-1.0}$~\ms, within $1$-$\sigma$ of the values reported in \citetalias{Yee2018} but with greater precision thanks to having covered a second orbit.

\vspace{6pt}

In summary, we presented new spectroscopic observations of HAT-P-11, extending the data baseline by six years. 
The precise RV measurements are consistent with the existence of a distant Jovian planet on an eccentric orbit.
Meanwhile, the S$_\mathrm{HK}$ activity indicator does not track the measured RVs on these timescales, indicating that stellar activity is only responsible for a fraction of the total observed RV variation.
We believe that the most parsimonious explanation for these data and the astrometric acceleration from Hipparcos and \textit{Gaia} is that HAT-P-11c is a true planet.
HAT-P-11 continues to be a benchmark system for studying planetary dynamics, atmospheres, and stellar activity, and the release of \textit{Gaia} epoch astrometry will likely unlock the full three-dimensional architecture of the system.

\begin{acknowledgments}
The authors wish to recognize and acknowledge the very significant cultural role and reverence that the summit of Maunakea has long had within the indigenous Hawaiian community. We are most fortunate to have the opportunity to conduct observations from this mountain.
\end{acknowledgments}

%

\vspace{5mm}
\facilities{Keck:I (HIRES)}


\software{radvel \citep{Radvel_Fulton18}
          }




\clearpage
\bibliography{hatp11c_revisited}{}
\bibliographystyle{aasjournal}

\end{document}